\documentclass[11pt,a4paper]{article}
\pdfoutput=1
\usepackage{jcappub}
\usepackage[utf8]{inputenc}
\usepackage{aas_macros}
\usepackage{amsfonts}
\usepackage{amsmath}
\usepackage{bm}
\usepackage{amssymb}
\usepackage{graphicx}
\usepackage[dvipsnames]{xcolor}
\usepackage{xfrac}
\usepackage{hyperref}
\usepackage{nameref}
\usepackage{cleveref}
\usepackage{multirow}
\usepackage{soul}
\usepackage{url}
\usepackage[thinlines]{easytable}
\crefformat{section}{\S#2#1#3}
\hypersetup{colorlinks=true,allcolors=teal}

\defcitealias{Arya_2023}{A23}

\providecommand{\keywords}[1]
{
  \small	
  \textbf{\textit{Keywords---}} #1
}

\newcommand{\Mpch}{\ensuremath{h^{-1}{\rm Mpc}}}

\newcommand{\be}{\begin{equation}}
\newcommand{\ee}{\end{equation}}

\title{\boldmath A modified lognormal approximation of the Lyman-$\alpha$ forest: comparison with full hydrodynamic simulations at $2\leq z\leq 2.7$}

\author[a,1]{B. Arya,\note{Corresponding author.}}
\author[b]{T. Roy Choudhury,}
\author[a]{A. Paranjape.}
\author[c]{and P. Gaikwad}

\affiliation[a]{Inter-University Centre for Astronomy \& Astrophysics,\\ Ganeshkhind, Post Bag 4, Pune 411007, India}
\affiliation[b]{National Centre for Radio Astrophysics, TIFR,\\Post Bag 3, Ganeshkhind, Pune 411007, India}
\affiliation[c]{Max-Planck-Institut für Astronomie,\\ Königstuhl 17, D-69117 Heidelberg, Germany}

\emailAdd{bharya@iucaa.in}
\emailAdd{tirth@ncra.tifr.res.in}
\emailAdd{aseem@iucaa.in}
\emailAdd{gaikwad@mpia-hd.mpg.de}

\abstract{Observations of the Lyman-$\alpha$ forest in distant quasar spectra with upcoming surveys are expected to provide significantly larger and higher-quality datasets. To interpret these datasets, it is imperative to develop efficient simulations. One such approach is based on the assumption that baryonic densities in the intergalactic medium (IGM) follow a lognormal distribution. 
We extend our earlier work to assess the robustness of the lognormal model of the Lyman-$\alpha$ forest in recovering the parameters characterizing IGM state, namely, the mean-density IGM temperature ($T_0$), the slope of the temperature-density relation ($\gamma$), and the hydrogen photoionization rate ($\Gamma_{12}$), by comparing with high-resolution Sherwood SPH simulations across the redshift range $2 \leq z \leq 2.7$. These parameters are estimated through a Markov Chain Monte Carlo (MCMC) technique, using the mean and power spectrum of the transmitted flux.
We find that the usual lognormal distribution of IGM densities cannot recover the parameters of the SPH simulations. This limitation arises from the fact that the SPH baryonic density distribution cannot be described by a simple lognormal form. To address this, we extend the model by scaling the linear density contrast by a parameter $\nu$. While the resulting baryonic density is still lognormal, the additional parameter gives us extra freedom in setting the variance of density fluctuations. With this extension, values of $T_0$ and $\gamma$ implied in the SPH simulations are recovered at $\sim 1-\sigma$ ($\lesssim$ 10\%) of the median (best-fit) values for most redshifts bins. However, this extended lognormal model cannot recover $\Gamma_{12}$ reliably, with the best-fit value discrepant by $\gtrsim 3-\sigma$ for $z > 2.2$. Despite this limitation in the recovery of $\Gamma_{12}$, whose origins we explain, we argue that the model remains useful for constraining cosmological parameters.}


\begin{document}
\label{firstpage}
\maketitle
\flushbottom

\section{Introduction}
\label{sec:intro}

The Lyman-$\alpha$ (Ly$\alpha$) forest observed in the spectra of distant quasi-stellar objects (QSOs) is a useful tracer for probing the underlying cosmic density field at relatively small scales \citep{Rauch_1998, Weinberg_2003, Meiksin_2009, McQuinn_2016}. The properties of the forest are sensitive to the thermal and ionization state of the intergalactic medium (IGM) and also to the underlying cosmological model. Hence it has been extensively used to constrain dark matter \citep{hansen_2002, Viel_2005, baur_2017, irsic_17b, bose_2019, garzilli_2019, palanque_2020, pedersen_2020, garzilli_2021, sarkar_2021}, cosmological \citep{1992AN....313..265L, 1998APS..APR.8WK05W, 2003MNRAS.342L..79S, 2003MNRAS.344..776M, Bird2023priya, tohfa2023forecast, Khan_2023} and astrophysical parameters \citep{2000bgfp.conf..455S, Bolton_2005, gaikwad2020, Gaikwad_2021}. With large surveys such as ongoing DESI \citep{ribera_2018, karacayli_2020, walther_2021, desi_2022, satya_2022} and upcoming WEAVE \citep{dalton_2012, weave_2022}, it will become possible to access a large number of QSO spectra, making it important to construct theoretical models and simulations which can be used for interpreting the data. In particular, efficient exploration of the unknown parameter space would require computationally efficient models of the IGM.

One approach for constructing such models is to make some approximation for the baryonic density field and use physical parameters to compute the Ly$\alpha$ optical depth, for examples of such models see \citep[][henceforth, A23]{Arya_2023} and references therein. More recently, there have been approaches based on effective field theory \citep{ivanov2023dont}, like those used in the study of large-scale structures. A different approach to efficient parameter space exploration is based on machine learning techniques, e.g., using generative neural networks in combination with low-resolution simulations to produce outputs equivalent to high-resolution hydrodynamical simulations \citep{jacobus2023reconstructing}. All of these approaches enable, in principle, a joint exploration of the cosmological and astrophysical parameter space. This is particularly important when considering parameters related to dark matter phenomenology (such as, e.g., the mass of a `warm' dark matter candidate) which lead to suppression of power at small scales, since such effects may also arise due to variations in the thermal history of the IGM.

In our earlier work \citetalias{Arya_2023}, we developed an end-to-end MCMC analysis method to constrain the astrophysical and cosmological parameters using the lognormal approximation. The lognormal model offers a quick and simplistic way of modelling the IGM. We tested the model in recovering the thermal and ionization parameters against Sherwood simulation, a smooth particle hydrodynamical (SPH) simulation, at $z \sim 2.5$. 
Building on \citetalias{Arya_2023}, in this paper we extend the work to other redshifts. We also improve the methodology used in previous work in several ways:
(i) We find that the lognormal model, in the usual implementation, is unable to match simultaneously the 1-point probability density function (PDF) and the power spectrum of the SPH simulations. To address this, we take a first step to modify the lognormal model and introduce an additional parameter, $\nu$, to scale the 1D baryonic density field.
(ii) Even with this additional parameter, we find that the lognormal model cannot match the flux probability distribution function (FPDF) particularly at high redshifts. We henceforth exclude FPDF from likelihood analysis and only use flux power spectrum (FPS) and mean transmitted flux ($\bar{F}$). This is a relatively common practice in the literature \citep{2013AAS...22132304R, Borde_2014, Nasir_2016, Villasenor_2023, Irsic_2023, cabayolgarcia2023neural}, and we leave a more detailed study of the behaviour of the FPDF to future work.
(iii) We use  a larger path length in SPH sightlines ($DX \sim 16$ compared to $\sim 6$ previously). This brings the SPH data closer to current high resolution observed data.
(iv) We also reduce uncertainty due to our lognormal model by taking a path length twice the size of SPH data (these two path lengths were equal in previous work).
(v) We calculate the Ly$\alpha$ optical depth using the density, velocity, and temperature fields provided by SPH instead using optical depth values directly. This ensures identical $\Gamma_{12}$ values and fitting function for recombination coefficient in both SPH and lognormal. 

The layout of the paper is as follows, in \cref{sec:prep}, we describe in detail the differences (as compared to \citetalias{Arya_2023}) in methodology for calculating flux statistics, covariance matrices, performing likelihood analysis as well as the introduction of additional parameter. In \cref{sec:result}, we present our results of recovering thermal and ionization histories. In \cref{sec:discuss}, we discuss the limitations of lognormal model as well as potential application, and we conclude in \cref{sec:conclude}.

\section{Simulations \& Method}
\label{sec:prep}
\noindent
In this section, we briefly describe the semi-numerical simulations of the Ly$\alpha$ forest based on the lognormal model, the Sherwood SPH simulations used for comparison and the procedure for parameter recovery using likelihood analysis and request the readers to refer to \citetalias{Arya_2023} for more details.
Throughout this work, we fix cosmological parameters for lognormal to Planck 2014 cosmology, the same being used in Sherwood simulations, \{$\Omega_m = 0.308$, $\Omega_{\Lambda} = 1 - \Omega_m$, $\Omega_b = 0.0482$ $h = 0.678$, $\sigma_8 = 0.829$, $n_s = 0.961$, $Y = 0.24$\}, consistent with the constraints from \citep{Planck_2014}. 

\subsection{Lognormal approximation}
In our framework, the linearly extrapolated power spectrum of DM density field, $P_{\mathrm{DM}}(k)$, is calculated for given set of cosmological parameters.\footnote{We use the CAMB transfer function \citep[\url{https://camb.readthedocs.io/en/latest/}]{camb} to calculate linear matter power spectrum, same as Sherwood simulations \citep{bolton+17-sherwood}.}
The 3D power spectrum of the baryonic density fluctuations at any given redshift $z$ is then given by\footnote{Unlike in some literature \citep{Kulkarni_2015, rorai_2017}, where smoothing is done on the Ly$\alpha$ transmitted flux, we use a more physical way by smoothing the DM density field itself.}
\begin{equation}
    P^{(3)}_{\mathrm{b}}(k, z) = D^2(z) P_{\mathrm{DM}}(k)~\mathrm{e}^{-2 x_{\mathrm{J}}^2(z) k^2}.
    \label{eq:Pb_PDM_gaussian}
\end{equation}
where $D(z)$ is the linear growth factor and $x_{\mathrm{J}}(z)$ is the Jeans length. The above relation is based on the assumption that the baryonic fluctuations follow the dark matter at large scales $k^{-1} \gg x_{\mathrm{J}}$ and are smoothed because of pressure forces at scales $k^{-1} \lesssim x_\mathrm{J}$. Since the Ly$\alpha$ forest probes the cosmic fields only along the lines of sight, it is sufficient to generate the line of sight baryonic density field $\delta_{\mathrm{b}}^L(x, z)$ and the corresponding line of sight component of the velocity fields $v_{\mathrm{b}}^L(x, z)$ only along one direction. We can obtain the 1D baryonic $(P^{(1)}_{\mathrm{b}}(k, z))$ and linear velocity $(P^{(1)}_{\mathrm{v}}(k, z))$ power spectra from 3D baryonic power spectra by
\begin{equation}
    P^{(1)}_{\mathrm{b}}(k, z) = \frac{1}{2\pi}\int_{|k|}^{\infty} dk^{\prime} k^{\prime} P^{(3)}_{\mathrm{b}}(k, z),
    \label{eq:Pb_1d}
\end{equation}and 
\begin{equation}
    P^{(1)}_{\mathrm{v}}(k, z) = \dot{a}^2(z) k^2 \frac{1}{2\pi}\int_{|k|}^{\infty} \frac{dk^{\prime}}{k^{\prime 3}} P^{(3)}_{\mathrm{b}}(k, z),
    \label{eq:Pv_1d}
\end{equation}
where $a$ is the scale factor $\dot{a}$ is given by the Friedman equation
\begin{equation}
    \dot{a}^2(z) = H_0^2\left[\Omega_{\mathrm{m}}(1 + z) + \Omega_{\mathrm{k}} + \frac{\Omega_{\Lambda}}{(1 + z)^2}\right],
\end{equation} with $\Omega_{\mathrm{k}} = 1 - \Omega_{\mathrm{m}} - \Omega_{\Lambda}$.
We then follow procedure given by \citep{csp01, Bi_1993} to generate density and velocity fields along line of sight using eqs.~\ref{eq:Pb_1d} and ~\ref{eq:Pv_1d}.

To account for the quasi-linear description of the density field, we employ the lognormal assumption and take the baryonic number density to be
\begin{equation}
     n_{\mathrm{b}}(x,z) = A~\mathrm{e}^{\delta^L_{\mathrm{b}}(x,z)}, \label{eq:LNapprox}
\end{equation}
where $A$ is a normalization constant fixed by setting the average value of $n_{\mathrm{b}}(x,z)$ to the mean baryonic density $\bar{n}_\mathrm{b}(z)$ at that redshift.

Given the density and velocity fields of baryons, one can compute the neutral hydrogen field assuming the gas to be in photoionization equilibrium,

\begin{equation}
     \alpha_A[T(x,z)]~n_{\mathrm{p}}(x,z)~n_{\mathrm{e}}(x,z) = n_{\mathrm{HI}}(x,z)~\Gamma_{12}(z)/(10^{12}~\mathrm{s}),
 \end{equation}
 where $\alpha_A(T)$ is the recombination coefficient at temperature $T$ (taken to be of A-type in this work, appropriate for the low-density IGM), $n_\mathrm{p}, n_\mathrm{e}$ are the number densities of protons and free electrons respectively and $\Gamma_{\mathrm{12}}$ is the hydrogen photoionization rate (in units of $10^{-12}$ s$^{-1}$ and assumed to be homogeneous). Assuming a fully ionized IGM, $n_\mathrm{p}, n_\mathrm{e}$ are given by,

\begin{equation}
     n_p(x,z) = \frac{4(1 - Y)}{4 - 3Y}n_{\textrm{b}}(x,z)\, ; \, n_e = \frac{4 - 2Y}{4 - 3Y}n_{\textrm{b}}(x,z)
\end{equation}where $Y (\sim 0.24)$ is helium weight fraction.
This requires specifying the IGM temperature at every point, which we do using a power-law temperature-density relation characterized by the temperature $T_0$ at the mean density and the slope $\gamma$, appropriate for the low-density IGM $T(x,z) = T_0(z)[1 + \delta^L_{\mathrm{b}}(x,z)]^{\gamma (z) - 1}$. We also need to assume the value of the photoionization rate. The Ly$\alpha$ optical depth is computed by accounting for thermal and natural broadening at each of these grid points $x_i$,
\begin{align}
    \tau(x_i, z) &= \frac{c I_{\alpha}}{\sqrt{\pi}} \sum_j \delta x \frac{n_{\mathrm{HI}}(x_j,z)}{b(x_j,z)[1+z(x_j)]}
    \notag \\
    &\times  V_{\alpha}\left(\frac{c[z(x_j)-z(x_i)]}{b(x_j,z)[1+z(x_i)]}+\frac{v^L_{\mathrm{b}}(x_j,z)}{b(x_j,z)}\right),
\end{align}
where $\delta x$ is the separation between the grid points (i.e., the grid size), $I_{\alpha} = 4.45\, \times 10^{-18}$ cm$^2$ is the Ly$\alpha$ absorption cross section and $V_{\alpha}(\Delta v / b)$ is the Voigt profile for the Ly$\alpha$ transition and 
\begin{equation}
    b(x,z) = \sqrt{\frac{2 k_{\mathrm{boltz}} T(x,z)}{m_{\mathrm{p}}}},
\end{equation}
where $m_\mathrm{p}$ is the proton mass. It then leads to the main observable, i.e., the normalized Ly$\alpha$ transmitted flux, $F(x_i, z) = e^{-\tau(x_i, z)}$. To mimic observational data, we also convolve $F(x_i, z)$ with Gaussian line spread function of full width at half maximum 7 km s$^{-1}$ as well as add random noise of SNR 50 per pixel. Our model at this stage is thus described by four free parameters, namely, \{$x_\mathrm{J}, T_0, \gamma, \Gamma_{12}$\}. 

Additionally, we introduce a new free parameter, $\nu$, which scales 1D baryonic density field according to
\begin{equation}
\delta^L_{\mathrm{b}}(x,z) \to \nu\,\delta^L_{\mathrm{b}}(x,z)\,.
\end{equation}

We emphasize that we perform this scaling \emph{before} exponentiating $\delta^L_{\mathrm{b}}$ in equation~\eqref{eq:LNapprox}. 
The inclusion of the parameter $\nu$ is an attempt to allow the model some freedom in setting the variance of baryonic density fluctuations (which are still distributed as a lognormal). As we already stated in \citetalias{Arya_2023}, the default model, which corresponds to the case $\nu=1$, is not a good description of baryonic properties. \emph{Our final model thus contains five free parameters:} $\{x_\mathrm{J}, T_0, \gamma, \Gamma_{12},\nu\}$.

\subsection{SPH simulation}
To test the validity of our model, we use publicly available Sherwood simulations suite \citep{bolton+17-sherwood} that were performed with a modified version of the cosmological smoothed particle hydrodynamics code P-Gadget-3, an extended version of publicly available GADGET-2 code \citep{Springel_2005}\footnote{\url{https://wwwmpa.mpa-garching.mpg.de/gadget/}}. The Sherwood suite consists of cosmological simulation boxes with volume
ranging from $10^3$ to $160^3$ $h^{-3}\,\textrm{cMpc}^3$ and contains number particles ranging from $2\times 512^3$ to $2 \times 2048^3$. The size and resolution of simulation box are suitable for
studying the small scale structures probed by Ly$\alpha$ forest.
The properties of Ly$\alpha$ forest from Sherwood simulation suite are well converged \citep{bolton+17-sherwood}. 
Similar to \citetalias{Arya_2023}, as the default, we choose a box of volume $40^3$ $h^{-3}\,\textrm{cMpc}^3$ containing $2 \times 2048^3$ particles.

\subsection{Skewer configuration and covariance matrices}
\begin{figure*}
\centering
\includegraphics[width=\textwidth]{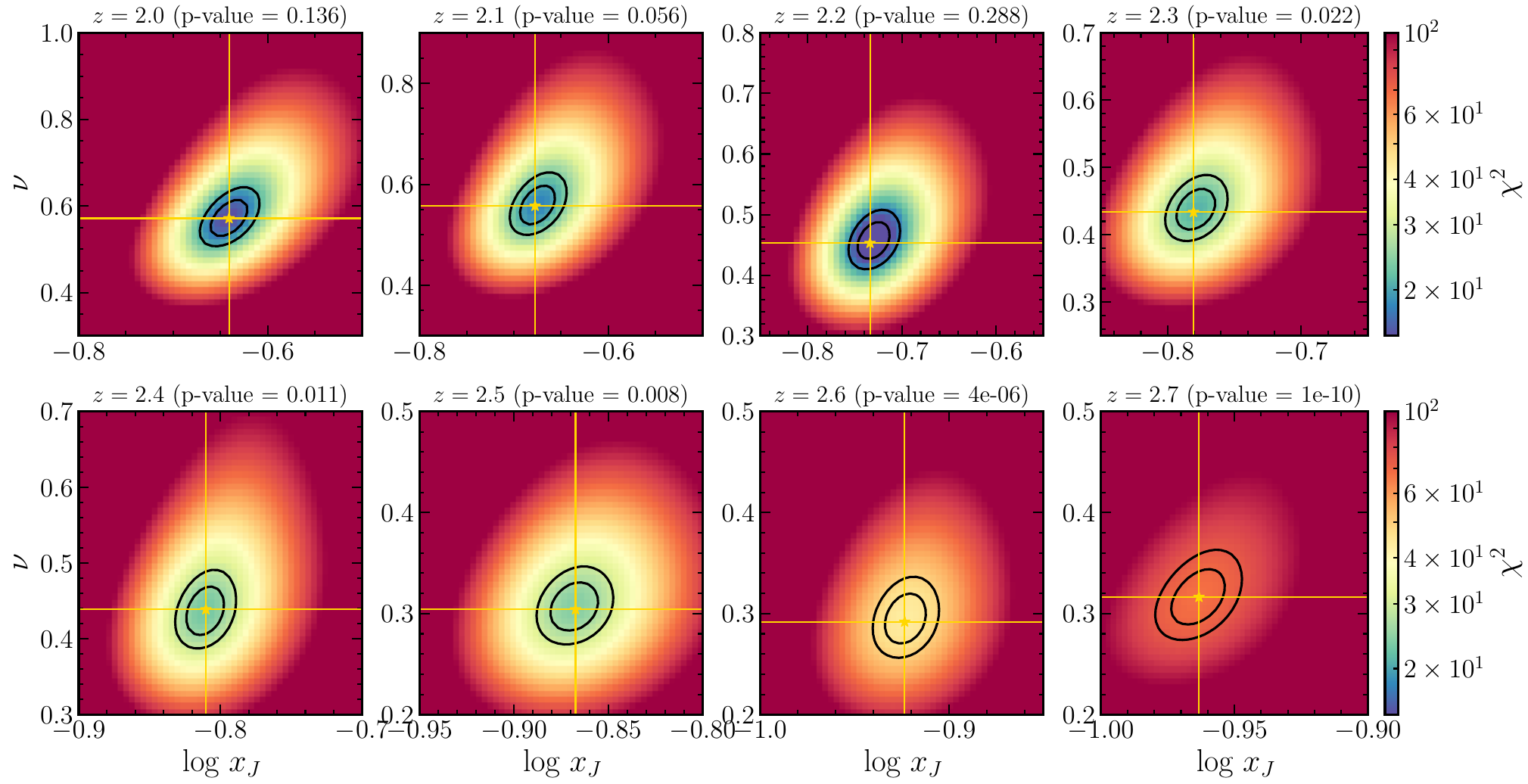}
\caption{$\chi^2$ colormap on log $x_{\textrm{J}}$ - $\nu$ grid with \{$T_0, \gamma, \Gamma_{12}$\} fixed to their true values for all 8 redshift bins. We get acceptable fits for $z \leq 2.5$. Black contours show 1 and 2-$\sigma$ confidence levels and gold stars show position of best-fit \{$x_{\mathrm{J}}$, $\nu$\}. p-values or probability-to-exceed (PTE) are mentioned along side redshift at top of each panel.}
\label{fig:2d_chi2}
\end{figure*}

\begin{figure*}
\centering
\includegraphics[width=\textwidth]{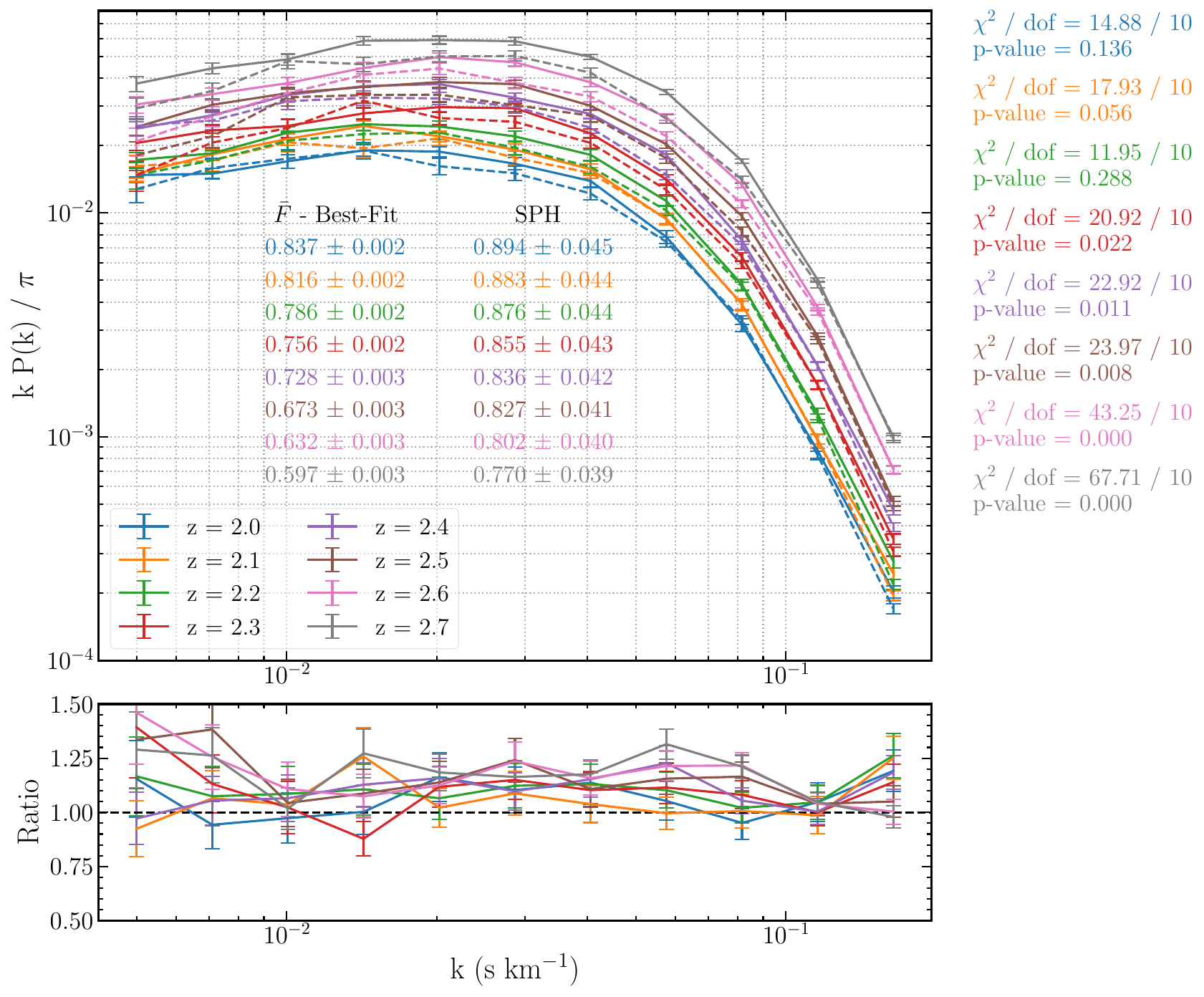}
\caption{Flux statistics for SPH data and best-fit parameters obtained from 2D $\chi^2$ analysis. Solid curves are best-fit lognormal and dashed curves are SPH.}
\label{fig:2d_chi2_stat}
\end{figure*}

The calculation of all relevant statistics including the mean flux, FPS and their covariances for both SPH and lognormal remains similar to \citetalias{Arya_2023} (see their equations~12-16). We calculate our "data points" by averaging statistics over 100 skewers picked randomly from a total 5000 available. The SPH covariance matrix is calculated using Jackknife resampling using the entire 50 (=5000/100) realizations. For lognormal covariance matrix, we generate 40000 skewers of same size of that of SPH. To reduce uncertainty from lognormal relative to SPH, the covariance matrix for lognormal is calculated by averaging statistics over 200 skewers. The covariance matrix is then calculated using 200 (=40000/200) realizations without Jackknife resampling. There are other key differences, which we list here:
\begin{itemize}
    \item In \citetalias{Arya_2023}, we had used a default path length of $DX \sim 6.2$ (equivalent to averaging over 40 skewers) at $z = 2.5$ in both lognormal and SPH. Additionally, we also considered a \textit{variation} where we averaged SPH and lognormal over 100 and 200 skewers respectively. Our reasoning was to bring the SPH dataset closer to current observational data and reduce uncertainties arising from lognormal relative to SPH. We consider this \textit{variation} as our \textit{default} configuration throughout this work. Please note that we have average SPH and lognormal over 100 and 200 skewers respectively at all redshifts and vary $DX$ accordingly.

    \item We exclude FPDF from the likelihood analysis as it gives poor fits at higher redshifts and only use \{$\bar{F}$ + FPS\} for the said purpose. In $\bar{F}$, we artificially scale the errors to 5\% of mean flux at every redshift in SPH data. The reason to scale the error on $\bar{F}$ is two-fold, (i) results from 2D $\chi^2$ analysis on log $x_{\mathrm{J}}$ - $\nu$ grid using unscaled error on $\bar{F}$ ($\sim 1\%$ of mean flux) and keeping other parameters fixed to their true values (similar to the results in Figs.~\ref{fig:2d_chi2} and \ref{fig:2d_chi2_stat}) showed minimum $\chi^2_{\nu}$ $\gtrsim$ 10 at all redshifts, implying that the model is unable to recover the true parameters. We were able to ascertain that the poor fits resulted from the inability of lognormal to match $\bar{F}$ within such small error. (ii) The observed mean flux typically has $\sim 5\%$ error arising due to systematic uncertainty in continuum placement \citep{Gaikwad_2021}.
\end{itemize}

We have run eight  Markov Chain Monte Carlo (MCMC)  chains using publicly available code \texttt{cobaya}\footnote{\url{https://cobaya.readthedocs.io/en/latest/sampler_mcmc.html}}\citep{Lewis_2002, Lewis_2013, Torrado_2021}, at $z =$ \{2, 2.1, 2.2, 2.3, 2.4, 2.5, 2.6, 2.7\}. To determine when a chain is converged, we use Gelman-Rubin statistics parameter, $R-1 < 0.05$ \citep{gelman_rubin, Lewis_2013}. The convergence for each chain takes $\sim 10$ days, using 32 cpus on a 2048 grid with a path length of $\sim 16$ at $z = 2.5$. 
All MCMC calculations were performed on the Pegasus cluster at IUCAA.\footnote{\url{http://hpc.iucaa.in/}}

\section{Results}
\label{sec:result}

In this section, we present the recovery of the free parameters of the lognormal by comparing with the SPH simulations. 

\begin{figure*}
\centering
\includegraphics[width=\textwidth]{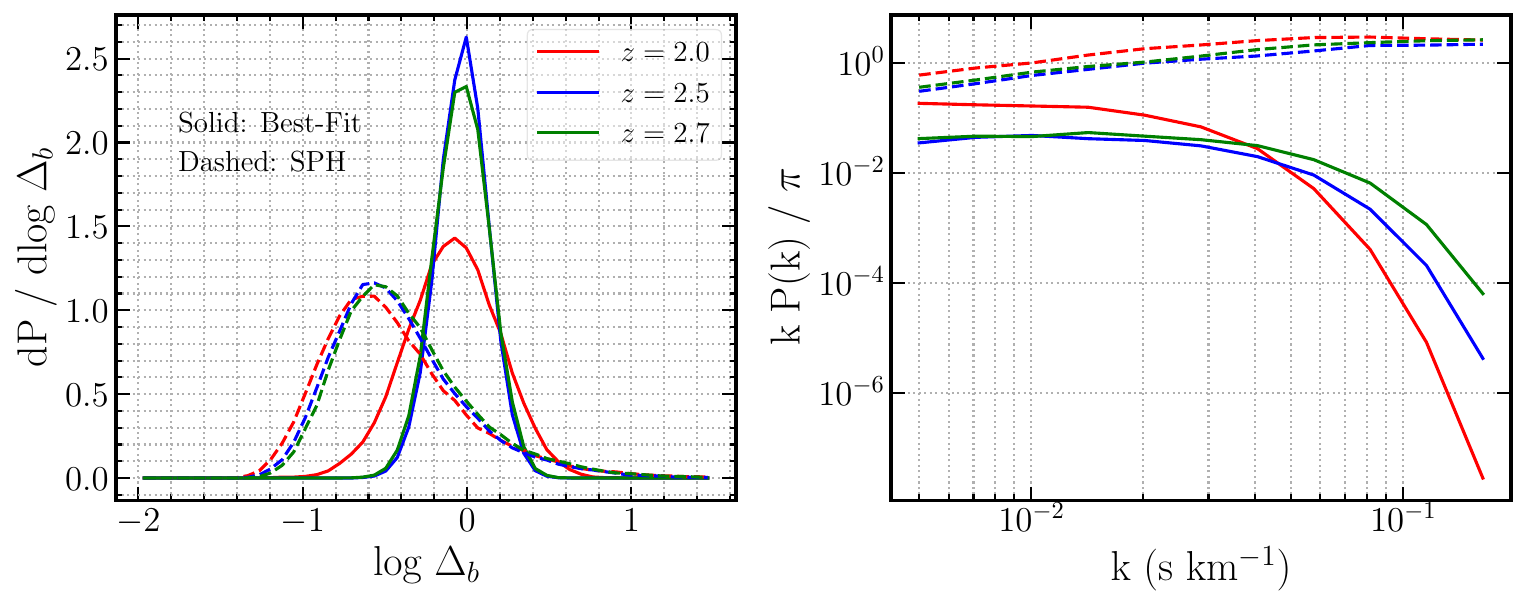}
\caption{1-point PDF of log $\Delta_{\rm b}$ (left) and power spectrum of $\Delta_{\rm b}$ (right) for best-fit \{$x_{\textrm{J}}$-$\nu$\}, keeping $\{T_0,\gamma,\Gamma_{12}\}$ fixed to their true values (solid curves), along with statistics for SPH (dashed curves).}
\label{fig:density_stat}
\end{figure*}

\begin{figure*}
\centering
\includegraphics[width=\textwidth]{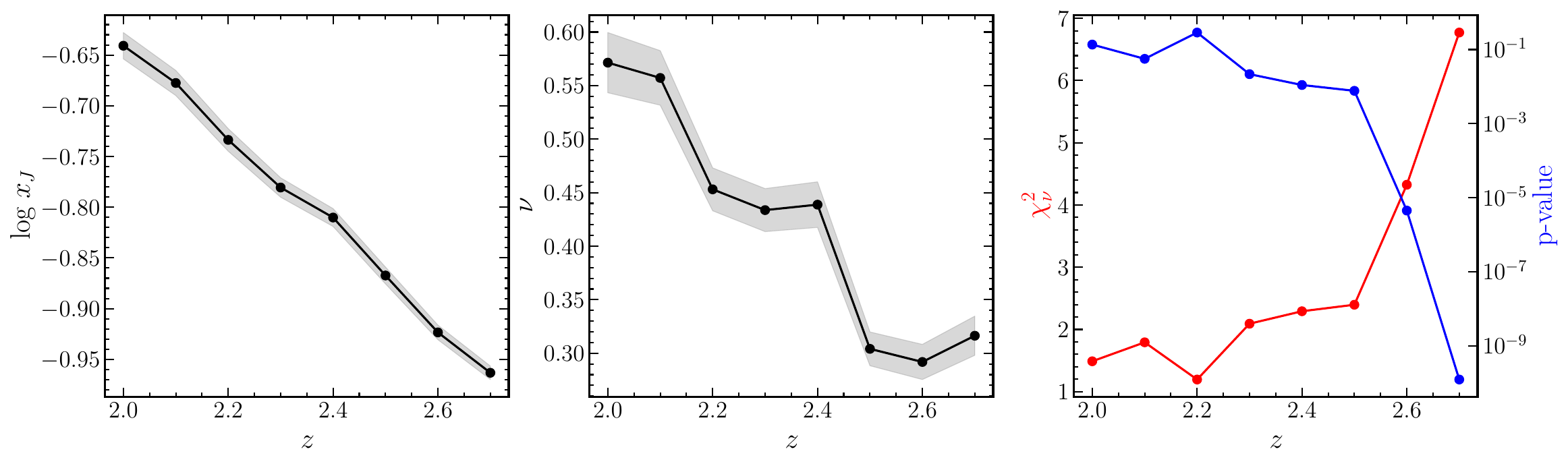}
\caption{First and second panels show redshift evolution of best-fit $x_{\textrm{J}}$ and $\nu$ from 2D $\chi^2$ analysis and true values of \{$T_0, \gamma, \Gamma_{12}$\} (black curves). Third panel shows minimum $\chi^2_{\nu}$ (red) and corresponding p-value (blue) at each redshift.}
\label{fig:2d_chi2_evol}
\end{figure*}

\subsection{2-parameter fit}

Before presenting MCMC results for our full 5-parameter model described in the previous sections, let us first try to understand what would be the typical value of the Jeans length $x_{\mathrm{J}}$ and $\nu$, parameters which do not have an obvious counterpart in the SPH simulations. Keeping this in mind, we do a simple $\chi^2$-minimization using a 2D grid in $\log$ $x_{\mathrm{J}}$ - $\nu$ and find the values of \{$x_{\mathrm{J}}$, $\nu$\} which best fit the simulation output. For the other three parameters, namely, $\Gamma_{12}, T_0, \gamma$, we use the values as in the SPH simulation. As mentioned earlier, we use the two flux statistics $\bar F$ and FPS for calculating the $\chi^2$. We repeat this exercise at all 8 redshifts.

Fig. \ref{fig:2d_chi2} shows the colormap plot of $\chi^2$ as a function of $x_{\mathrm{J}}$ and $\nu$ for all 8 redshifts. Fig.~\ref{fig:2d_chi2_stat} shows  the corresponding best-fit lognormal flux statistics compared with the input SPH. Fig.~\ref{fig:2d_chi2_evol} summarises the redshift evolution of best-fit $x_{\mathrm{J}}$ and $\nu$. At every redshift, we see clear minima in Fig. \ref{fig:2d_chi2} which provides monotonically decreasing (\textit{w.r.t.} redshift) best-fit values of $x_{\textrm{J}}$ from 0.23 $\Mpch$ at $z = 2$ to 0.11 $\Mpch$ at $z = 2.7$ in Fig.~\ref{fig:2d_chi2_evol}. This value is of the same order as the one obtained by assuming Ly$\alpha$ absorbers to be in hydrostatic equilibrium at a temperature $\sim 10^4$K \citep{schaye_2001}. The best-fit values of $\nu$ do not show such simple monotonic trend, although still decrease as we go from lowest to highest redshift. The minimum reduced $\chi^2$, $\chi^2_{\nu,\textrm{min}}$ and p-values for $2 \leq z \leq 2.5$ are \{1.5, 1.8, 1.2, 2.1, 2.3, 2.4\} and \{0.136, 0.056, 0.288, 0.022, 0.011, 0.008\} respectively, which implies that the fit is just acceptable. The fits for $z = 2.6$ and 2.7 are however, quite poor with $\chi^2_{\nu,\textrm{min}}$ and p-values being \{4.3, 6.8\} and \{4 $\times$ 10$^{-6}$, 1 $\times$ 10$^{-10}$\} respectively (Fig.~\ref{fig:2d_chi2_stat}). This poor fit implies that the lognormal cannot recover the SPH statistics when the three parameters $T_0, \gamma, \Gamma_{12}$ are fixed to the SPH values. 

To understand the possible reasons for this failure at higher redshifts, in Fig.~\ref{fig:density_stat} we show the 1-point PDF \emph{(left panel)} and power spectrum \emph{(right panel)} of the baryonic number density fluctuation $\Delta_{\rm b}=n_{\rm b}/\bar n_{\rm b}$ calculated in the lognormal approximation at three redshifts, using the SPH values for $\{T_0,\gamma,\Gamma_{12}\}$ and the best fit values of $\{x_J,\nu\}$ from the 2-d analysis described above. 
To our knowledge this is the first comparison of $\Delta_{\rm b}$ statistics between SPH and a lognormal model weakly adjusted to fit the $\bar F$ and FPS statistics. It is quite apparent that neither the PDF nor the power spectrum of $\Delta_{\rm b}$ the lognormal model agree well with the SPH quantities at any redshift. Nevertheless, at least at low redshifts $z\simeq2.0$, the model produces a completely acceptable fit to the FPS and $\bar F$ measurements. E.g., it is striking that the orders of magnitude difference between the power spectra at $z=2$ (red solid and dashed curves in the \emph{right panel} of Fig.~\ref{fig:density_stat}) is consistent with the excellent match in FPS and $\bar F$ seen in the blue curves in Fig.~\ref{fig:2d_chi2_stat}. Simultaneously, the mean of $\log\Delta_{\rm b}$ is clearly higher in this best fitting lognormal model than in the SPH. The power spectrum conundrum might be partially attributed to the fact that the FPS is a highly smoothed version (due to the Voigt kernel) of a highly nonlinear transform of $\Delta_{\rm b}$, such that the scales relevant for calculating the FPS from $\Delta_{\rm b}$ are restricted to small $k$ (indeed, the conventional wisdom is that these are `quasi-linear' scales). The mismatch in PDF of $\Delta_{\rm b}$, however, leaves a distinct imprint in the comparison of $\bar F$ values: $\bar F$ is systematically smaller in the best fitting lognormal model than in the SPH, with the difference becoming more pronounced (and more statistically significant) at larger redshifts (where the quality of the overall fit also degrades). This is can be qualitatively understood in the context of the mismatch in the PDF of $\Delta_{\rm b}$; the overestimate of volume occupied by mildly overdense regions $\Delta_{\rm b}\sim1$-$3$ in the lognormal directly implies an underestimate of overall flux $F={\rm e}^{-\tau}$ at fixed photoionization rate, assuming that $\tau$ is approximately monotonic with $\Delta_{\rm b}$ and that only mildly overdense regions contribute to Ly$\alpha$ flux.

These arguments are not complete by any means, however. The $z=2$ results especially indicate that a substantial role might be played by higher order statistics (non-Gaussianities) of the baryonic log-density field, which the lognormal model simply sets to zero. The impact of these non-Gaussianities on the FPS and $\bar F$ has not been explored in the literature, to our knowledge. We leave such a study to future work. For now, we simply note that, despite the degrees of freedom provided by $x_J$ and $\nu$, the lognormal model constrained by Ly$\alpha$ flux statistics is unable to match the 1-point and 2-point statistics of the baryonic density fields. This suggests that, when varying all 5 model parameters, \emph{at least} one of the parameters $\{T_0,\gamma,\Gamma_{12}\}$ is likely not to be recovered with good accuracy. The arguments above would indicate that this parameter is likely to be the photoionization rate $\Gamma_{12}$, due to its impact on $\bar F$. 

Keeping this in mind, we proceed as in \citetalias{Arya_2023} to vary all the model parameters simultaneously in the next section. We will return to a discussion of the quality of parameter recovery in section~\ref{sec:discuss}.

\subsection{5-parameter fit}

\begin{figure*}
\centering
\includegraphics[width=\textwidth]{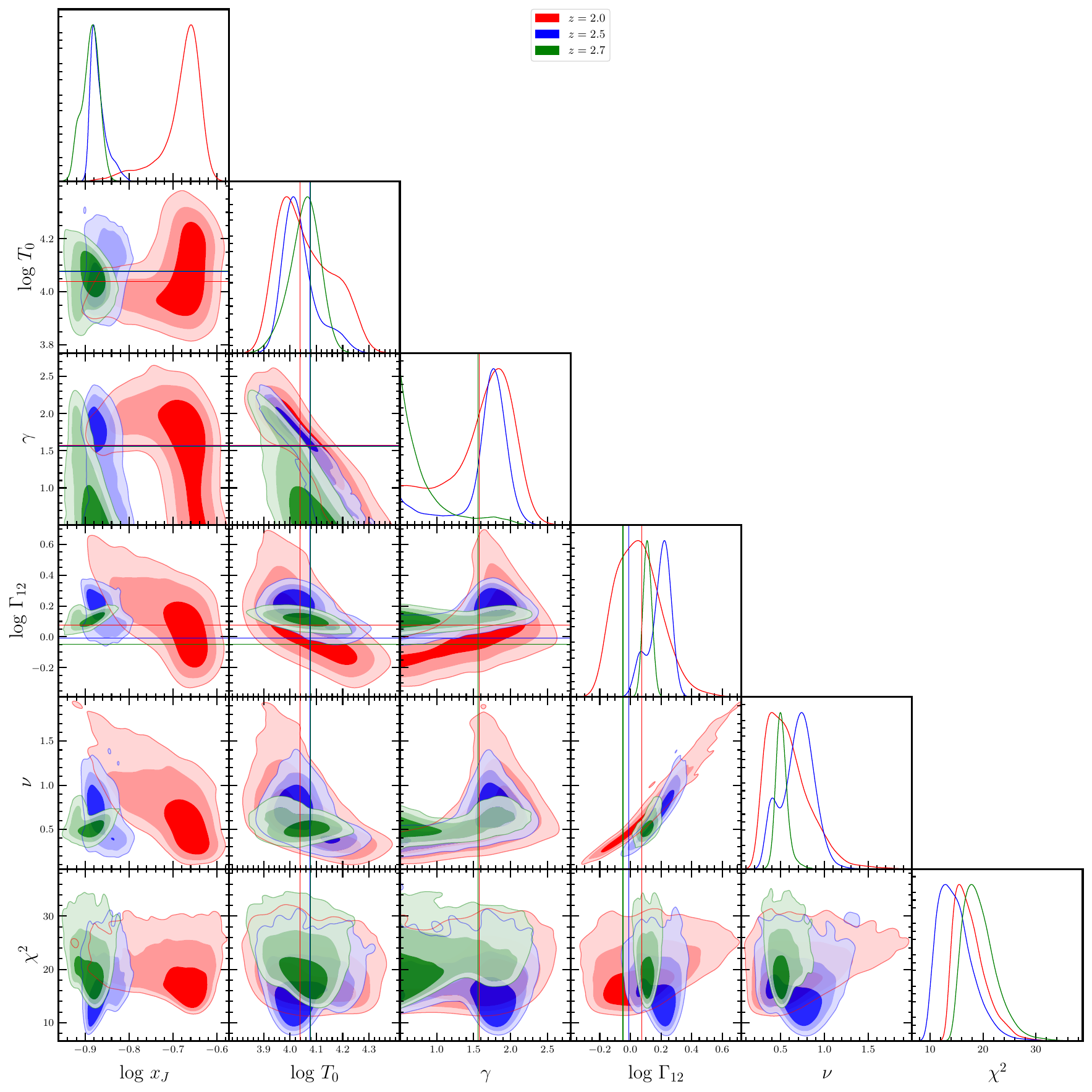}
\caption{1, 2, and 3$-\sigma$ contours for three redshifts, $z = 2, \, 2.5, \, \textrm{and} \, 2.7$. Colour coded horizontal and vertical lines show true values of parameters.}
\label{fig:corner_main}
\end{figure*}

\begin{figure*}
\centering
\includegraphics[width=\textwidth]{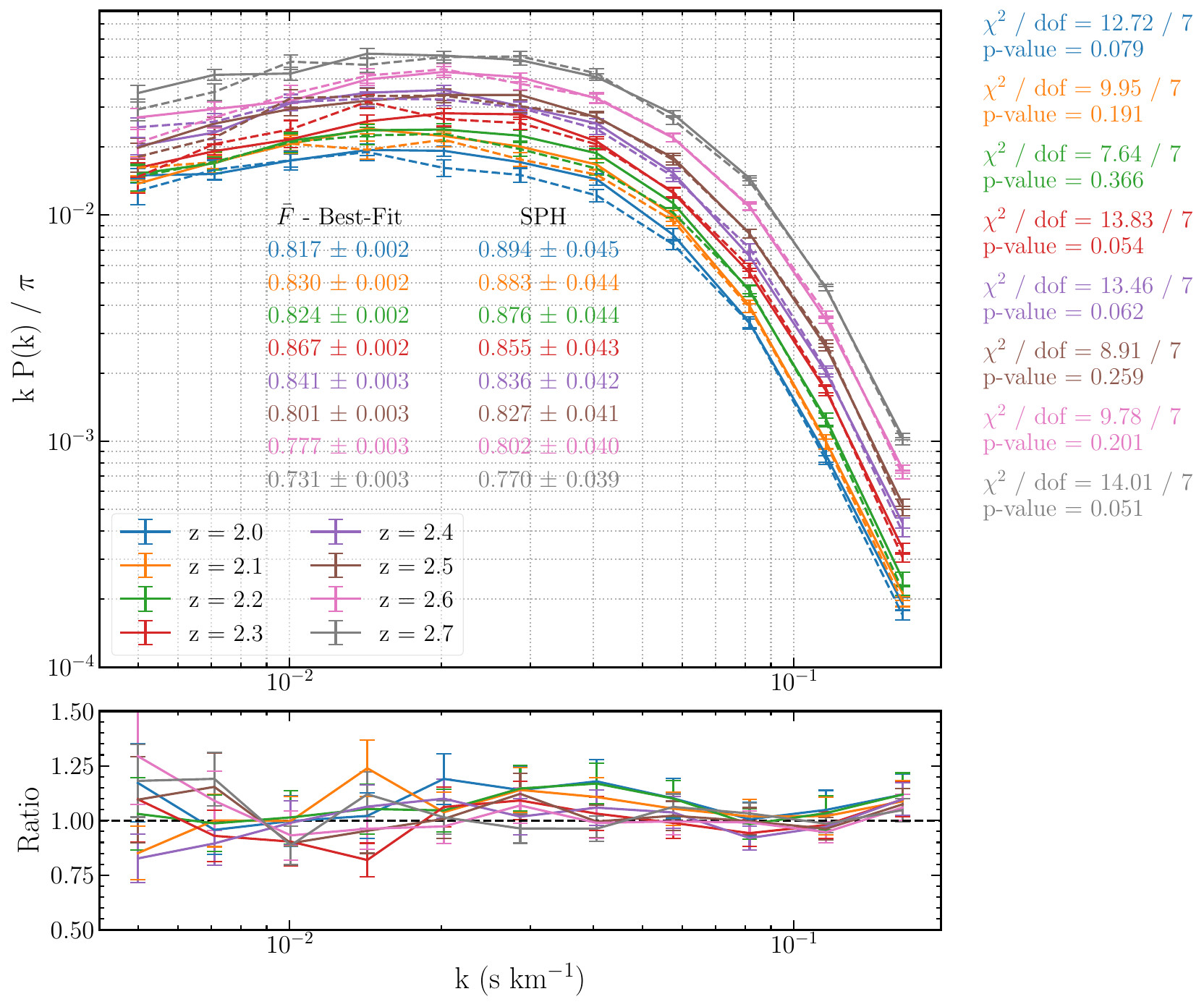}
\caption{Flux statistics for best-fit model and SPH data for all the redshifts.}
\label{fig:stat_main}
\end{figure*}

\begin{figure*}
\centering
\includegraphics[width=\textwidth]{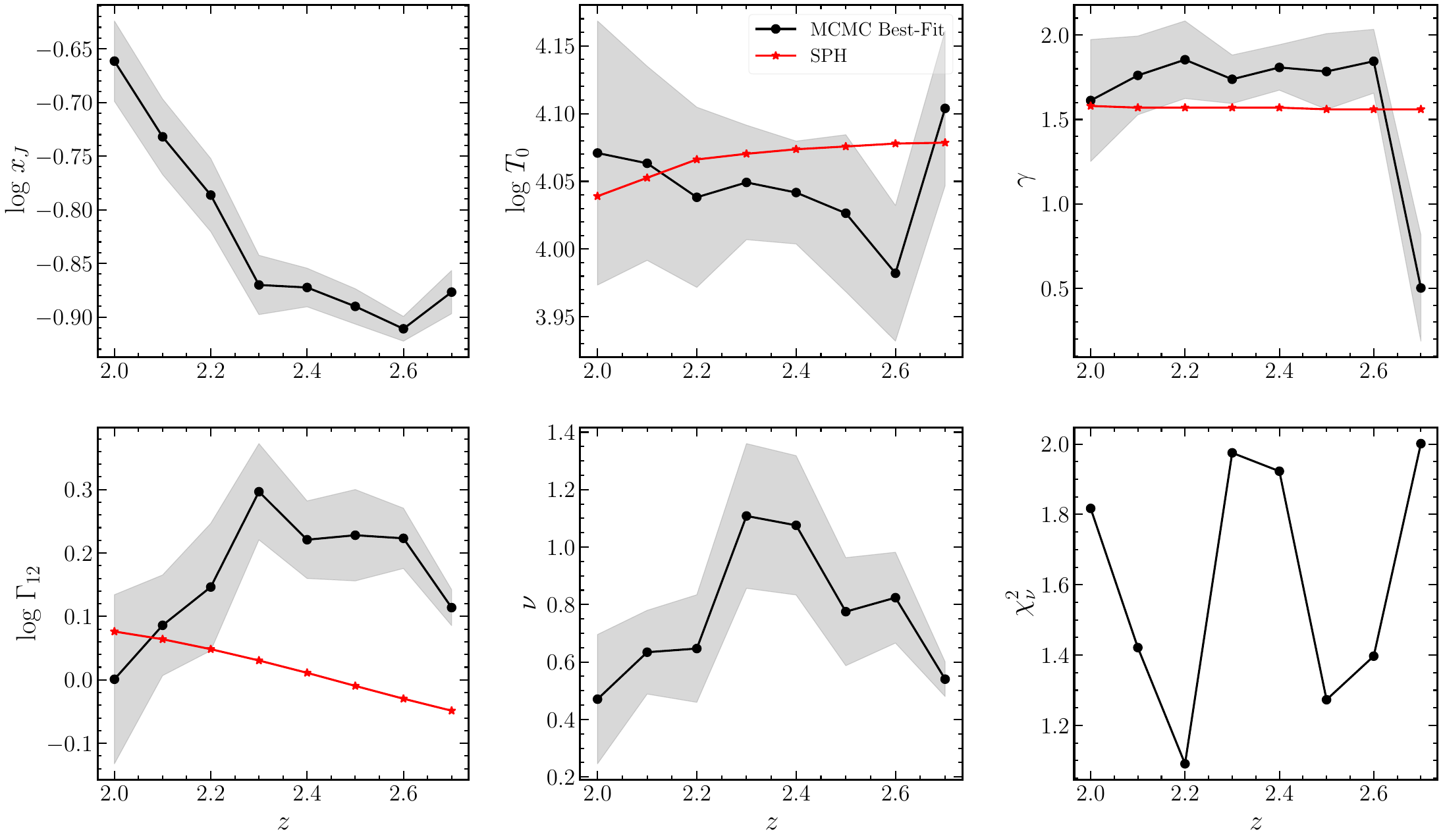}
\caption{Redshift evolution of parameters and reduced $\chi^2$ shown with black circles. Gray shaded regions 16 and 84 percentiles from MCMC chains. Red triangles are true values of parameters in SPH.}
\label{fig:param_evol}
\end{figure*}

In this section, we explore the case where all the free parameters of the lognormal are allowed to vary. Table \ref{table:priors} lists the priors on parameters, $\{\textrm{log}\, x_{\textrm{J}},\, \textrm{log}\, T_0, \, \gamma,\, \textrm{log}\, \Gamma_{12},\, \nu\}$. While we have used simple, flat, and wide priors on four parameters, $\{\textrm{log}\, T_0, \, \gamma,\, \textrm{log}\, \Gamma_{12},\, \nu\}$, we have imposed a more physically motivated prior on log $x_{\textrm{J}}$ by calculating the lower limit on the prior, $x_{\textrm{J,th}}$, using equation
\begin{equation}
    x_{\textrm{J,th}} = \frac{1}{H_0}\left[\frac{2\gamma_{\textrm{s}} k_{\textrm{B}}T_{0,\textrm{s}}}{3\mu m_{\textrm{p}}\Omega_{\textrm{m}}(1+z)}\right]^{1/2}
    \label{eq:xJ_th}
\end{equation}
where $T_{0,s}$ and $\gamma_{s}$ are values of $T_0$ and $\gamma$ sampled in the MCMC chain respectively. For reference, the values of log $x_{\textrm{J,th}}$ at redshifts \{2.0, 2.1, 2.2, 2.3, 2.4, 2.5, 2.6, 2.7\} for corresponding true values of $T_0$ and $\gamma$ are \{-0.877, -0.879, -0.879, -0.883, -0.895, -0.900, -0.905\} respectively.
The motivation behind imposing the limit from eq.\ref{eq:xJ_th} is that lognormal model tends to favor unphysically small values of $x_{\textrm{J}}$, $\sim 0.01\, \Mpch$ at high redshifts.
The upper limit on the prior (for log $x_{\textrm{J}}$) is fixed to 0.5.

\begin{table}
\centering
\begin{tabular}{||c c||} 
\hline
Parameter & Prior \\ [0.5ex]
\hline\hline
log $x_{\textrm{J}}$ & [log $x_{\textrm{J,th}}$, 0.5]\\ 
\hline
log $T_0$ & [2.5, 5.5]\\ 
\hline
$\gamma$ & [0.5, 5]\\
\hline
log $\Gamma_{12}$ & [-2, 2]\\
\hline
$\nu$ & [0.05, 2]\\
\hline
\end{tabular}
\caption{Priors on parameters, $\{\textrm{log}\, x_{\textrm{J}},\,\textrm{log}\, T_0, \, \gamma,\, \textrm{log}\, \Gamma_{12},\, \nu\}$, where log $x_{\textrm{J,th}}$ is calculated using eq.\ref{eq:xJ_th}. See text  for a discussion of the parameter $x_{\rm J}$.}
\label{table:priors}
\end{table}

The true values (i.e., the values used in or obtained from the SPH simulations) alongwith best-fit and median of the parameters at each redshift are reported in Table \ref{table:true_val_table}. In figs. \ref{fig:corner_main}, \ref{fig:stat_main}, and \ref{fig:param_evol} we show the contour plots (68.3, 95.4, 99.7 percentiles) for three redshifts, $z = 2, \, 2.5, \, \textrm{and} \, 2.7$, obtained from MCMC run, corresponding best-fit and SPH flux statistics, and the evolution of best-fit values of parameters with redshift respectively. From Fig. \ref{fig:param_evol}, it is evident that the lognormal model does a decent job at recovering $T_0$ at all redshifts with the true values being within 1$-\sigma$ from the median. $\gamma$, although, is recovered reasonably well at all redshifts except $z = 2.7$, has a tendency to favor bimodality with an "inverted" $T-\Delta_{\textrm{b}}$ relation. At $z = 2.7$, $\gamma$ has completely hit the prior. Another limitation of lognormal model is in estimating $\Gamma_{12}$. In fig.\ref{fig:param_evol}, we can see (i) $\Gamma_{12}$ is recovered within 1$-\sigma$ only till $z = 2.2$ and (ii) shape of evolution of $\Gamma_{12}$ \textit{w..r.t} redshift from lognormal is in complete disagreement with that of SPH. From both figs. \ref{fig:corner_main} and \ref{fig:param_evol}, we also observe degeneracies between $T_0 - \gamma$ and $\Gamma_{12} - \nu$. These degeneracies may significantly affect parameter estimates. E.g., underestimating value of $\gamma$ at $z = 2.7$ has overestimated $T_0$. Although in this particular case the strong anti-correlation has made the estimate of $T_0$ better. On the other hand, we see a strong positive correlation in $\Gamma_{12}$ and $\nu$ where shapes of redshift evolution curves of both parameters are remarkably similar.

Thus, as anticipated earlier, the lognormal model cannot simultaneously recover the true values of all the parameters $\{T_0,\gamma,\Gamma_{12}\}$ and it is indeed $\Gamma_{12}$ that is affected the most. We discuss this aspect of the model in more detail in the next section.

\section{Discussion}
\label{sec:discuss}

In this section we present a qualitative argument to understand the failure of the lognormal model seen above, followed by a discussion of potential cosmological applications of the model as it stands.

\subsection{Recovery of $\Gamma_{12}$}
The fact that the failure of the lognormal model in parameter recovery mostly manifests in the poor recovery of $\Gamma_{12}$ can be understood as follows. $\Gamma_{12}$ appears in the model entirely as a multiplicative factor in calculating the neutral hydrogen number density $n_{\rm HI}\propto\Gamma_{12}^{-1}$. This subsequently propagates to the optical depth $\tau(z)\sim\int{\rm d}z^\prime\,n_{\rm HI}(z^\prime)K(z,z^\prime)$, where the nature of the kernel $K$ is irrelevant for this argument since it does not involve $\Gamma_{12}$ (so we can also simply approximate $\tau\sim n_{\rm HI}$ as far as $\Gamma_{12}$ and $\nu$ are concerned). Thus $\tau\propto\Gamma_{12}^{-1}$ due to the dependence on $n_{\rm HI}$. However, since $n_{\rm HI}\propto n_{\rm b}^\beta$ where $\beta$ is order unity, the \emph{fluctuations} in $\tau$ (at least at large scales) scale proportionally to the parameter $\nu$. Thus, fluctuations in $\tau$ have the overall approximate dependence $\delta\tau\propto \nu\,\Gamma_{12}^{-1}$, making $\nu$ and $\Gamma_{12}$ highly degenerate when determining the power spectrum of the flux $F={\rm e}^{-\tau}$, and explaining the strong positive degeneracy mentioned above.

We further note that using the true value of $\Gamma_{12}$ leads to rather small values of $\nu$ in the previous 2-d analysis (Fig.~\ref{fig:2d_chi2}). Simultaneously, the best fitting model leaves behind a mean flux $\bar F = \langle\,{\rm e}^{-\tau}\,\rangle$ that is significantly smaller than the SPH, and an FPS that is significantly larger, especially at higher redshifts (see Fig.~\ref{fig:2d_chi2_stat}).   This is consistent with our earlier discussion regarding the fact that the lognormal approximation does not simultaneously describe the 1-pont and 2-point functions of the baryonic density field.

Once $\Gamma_{12}$ is also allowed to vary, the model can accommodate a larger value of $\bar F$ and smaller FPS by increasing both $\nu$ and $\Gamma_{12}$. To understand why, it is again useful to approximate $\Delta_{\rm b}\simeq1+\delta^L_{\rm b}$, so that $\tau\sim n_{\rm HI}\sim n_b^\beta/\Gamma_{12}\sim\bar n_b^\beta(1+\beta\delta_{\rm b}^L)/\Gamma_{12}$. The key point to note here is that $\Gamma_{12}$ multiplies both the mean and fluctuations of $\tau$, while $\nu$ appears only in the fluctuations, through $\delta_{\rm b}^L$. We can then approximate
\begin{equation}
\tau=\bar\tau+\delta\tau\sim C_1/\Gamma_{12}+C_2(\nu/\Gamma_{12})\hat g\,, 
\end{equation}
where $C_1$ and $C_2$ don't depend on $\nu$ or $\Gamma_{12}$ and $\hat g$ is a Gaussian distributed random variable with zero mean and unit variance. As compared to $C_1$, the quantity $C_2$ involves the amplitude of density fluctuations, so we expect $C_2\ll C_1$.
This gives us
\begin{align}
F &= {\rm e}^{-\tau} \sim {\rm e}^{-C_1/\Gamma_{12} - C_2(\nu/\Gamma_{12})\hat g} \,,\notag\\
\bar F &\simeq {\rm e}^{-\bar\tau + \langle\delta\tau^2\rangle/2} \sim {\rm e}^{-C_1/\Gamma_{12} + C_2^2(\nu/\Gamma_12)^2/2} \sim {\rm e}^{-C_1/\Gamma_{12}}\,,\notag\\
\delta_F &\equiv F/\bar F-1\simeq {\rm e}^{-\delta\tau-\langle\delta\tau^2\rangle/2}-1\simeq-\delta\tau\,,\notag\\
P_F &\sim \langle\delta_F^2\rangle \sim \langle\delta\tau^2\rangle \sim C_2^2(\nu/\Gamma_{12})^2\,.
\end{align}

We now consider the fact that, with $\nu=1$ and $\Gamma_{12}=\Gamma_{12}^{\rm (SPH)}$, the lognormal model produces a FPS ($\bar F$) that substantially overestimates (underestimates) that from SPH. If $\Gamma_{12}$ is fixed, the only available degree of freedom is $\nu$, which is driven to small values that decrease the FPS (since $P_F\propto\nu^2$) and \emph{also decrease} $\bar F$ (since $\bar F \sim {\rm e}^{\#\nu^2}$ when only $\nu$ is varied). Note that arbitrarily small values of $\nu\to0$ are strongly disfavoured by the FPS, which would be driven to zero in this case. 
The variation of $\Gamma_{12}$ in the full analysis now becomes significant; increasing $\Gamma_{12}$ \emph{increases} the value of $\bar F$ in the the lognormal model, since $\bar F\sim{\rm e}^{-C_1/\Gamma_{12}}$ (we argued above that the contribution of the term involving $C_2$ will be subdominant compared to that involving $C_1$). The interplay between $\nu$ and $\Gamma_{12}$ then balances out to match the FPS. The remaining leeway in achieving this balance shows up as the degeneracy between $\nu$ and $\Gamma_{12}$ we commented on earlier. Consistently with this argument, we have found that \emph{excluding} $\bar F$ from the inference leads to an essentially unbroken degeneracy between $\nu$ and $\Gamma_{12}$, which also highlights the important role played by the $\bar F$ constraint in our analysis. 

Evidently, the upshot is that a reasonably good fit to the FPS and $\bar F$ from SPH can, in fact, be achieved by the lognormal model, at the cost of significantly \emph{overestimating} $\Gamma_{12}$ (Fig.~\ref{fig:param_evol}). This might also be understood in a simpler manner, using the comparison of the 1-point PDF of $\Delta_{\rm b}$ between SPH and lognormal in Fig.~\ref{fig:density_stat}. There, we saw that the lognormal model with any reasonable value of $\nu$ significantly over-predicts the number of pixels having $\Delta_{\rm b}\sim1$-$3$, which is where one expects the Ly$\alpha$ forest to arise from. To compensate for this overestimate of potential neutral absorber systems, the model must increase the photoionization rate. As we mentioned earlier, this simple argument also indicates that the eventual resolution of the fact that the lognormal model fails to accurately reproduce Ly$\alpha$ flux statistics is likely related to understanding the higher moments of the baryonic log-density field (which the lognormal model treats as Gaussian distributed). Simple fixes such as the inclusion of the scaling parameter $\nu$ clearly have their limitations. Addressing this challenge while retaining the simplicity of this semi-analytical model is the subject of work in progress.

\begin{table*}
\centering
\scalebox{0.7}{
\begin{tabular}{||c c c c c c||} 
\hline
Redshift & log $x_{\textrm{J}}$ ($\Mpch$) & log $T_0$ (K) & $\gamma$ & log $\Gamma_{12}$ (10$^{-12}$ s$^{-1}$) & $\nu$ \\[0.8ex]
\hline\hline
2.0 & - / -0.66(-0.67$^{+0.03}_{-0.05}$) & 4.04 / 4.07(4.03$^{+0.12}_{-0.08}$)  & 1.58 / 1.61(1.75$^{+0.28}_{-0.44}$) & 0.08 / 0.00(0.07$^{+0.14}_{-0.13}$) & - / 0.47(0.58$^{+0.26}_{-0.19}$) \\[0.8ex]
\hline
2.1 & - / -0.73(-0.74$^{+0.03}_{-0.04}$) & 4.05 / 4.06(4.06$^{+0.08}_{-0.06}$) & 1.57 / 1.76(1.77$^{+0.23}_{-0.24}$) & 0.07 / 0.09(0.10$^{+0.08}_{-0.08}$) & - / 0.63(0.68$^{+0.16}_{-0.13}$) \\[0.8ex]
\hline
2.2 & - / -0.79(-0.79$^{+0.03}_{-0.04}$) & 4.07 / 4.04(4.05$^{+0.08}_{-0.05}$) & 1.57 / 1.85(1.77$^{+0.21}_{-0.25}$) & 0.05 / 0.15(0.15$^{+0.10}_{-0.10}$) & - / 0.65(0.66$^{+0.20}_{-0.17}$) \\[0.8ex]
\hline
2.3 & - / -0.87(-0.84$^{+0.03}_{-0.02}$) & 4.07 / 4.05(4.03$^{+0.05}_{-0.04}$) & 1.57 / 1.74(1.78$^{+0.16}_{-0.14}$) & 0.03 / 0.30(0.25$^{+0.07}_{-0.08}$) & - / 1.11(0.97$^{+0.26}_{-0.25}$) \\[0.8ex]
\hline
2.4 & - / -0.87(-0.86$^{+0.02}_{-0.01}$) & 4.07 / 4.04(4.03$^{+0.04}_{-0.04}$) & 1.56 / 1.81(1.83$^{+0.14}_{-0.13}$) & 0.01 / 0.22(0.19$^{+0.05}_{-0.07}$) & - / 1.08(0.96$^{+0.24}_{-0.25}$) \\[0.8ex]
\hline
2.5 & - / -0.89(-0.87$^{+0.02}_{-0.01}$) & 4.08 / 4.03(4.02$^{+0.07}_{-0.05}$) & 1.56 / 1.78(1.74$^{+0.18}_{-0.27}$) & -0.01 / 0.23(0.20$^{+0.06}_{-0.09}$) & - / 0.78(0.70$^{+0.17}_{-0.21}$) \\[0.8ex]
\hline
2.6 & - / -0.91(-0.90$^{+0.01}_{-0.01}$) & 4.08 / 3.98(3.99$^{+0.06}_{-0.05}$) & 1.56 / 1.85(1.79$^{+0.17}_{-0.21}$) & -0.03 / 0.22(0.20$^{+0.04}_{-0.06}$) & - / 0.82(0.74$^{+0.14}_{-0.17}$) \\[0.8ex]
\hline
2.7 & - / -0.88(-0.88$^{+0.02}_{-0.02}$) & 4.08 / 4.10(4.05$^{+0.05}_{-0.06}$) & 1.56 / 0.50(0.77$^{+0.44}_{-0.19}$) & -0.05 / 0.11(0.11$^{+0.03}_{-0.03}$) & - / 0.54(0.50$^{+0.07}_{-0.06}$) \\[0.8ex]
\hline
\end{tabular}}
\caption{True / best-fit(median) values for the parameters explored in MCMC run. Please see that $x_{\textrm{J}}$ and $\nu$ do not have any "true" values.}
\label{table:true_val_table}
\end{table*}

\subsection{Potential application}
\noindent
It is evident from the preceding discussion that $\Gamma_{12}$ cannot be reliably estimated using the present form of the lognormal model. Nevertheless, given its success in recovering the remaining IGM parameters, it is interesting to ask whether this model can still be used for \emph{cosmological} parameter inference. This might be the case if, e.g., the cosmological parameter directions are largely independent of $\Gamma_{12}$ in the space defined by the likelihood function.

\begin{figure*}
\centering
\includegraphics[width=\textwidth]{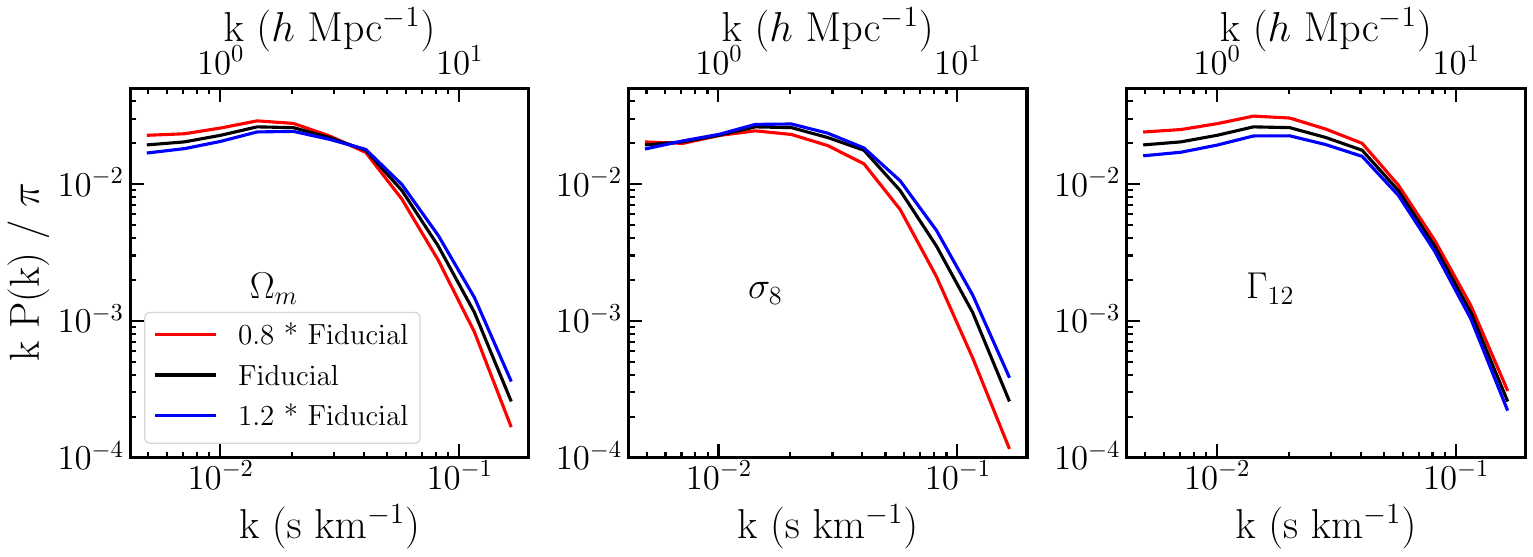}
\caption{Effect of parameters, \{$\Omega_{\textrm{m}}$, $\sigma_8$, $\Gamma_{12}$\} on flux power spectrum produced by lognormal model at $z = 2$. Black curves show FPS for a set of fiducial parameters while in red (blue) curves, we decrease (increase) parameters by 20\% one by one, keeping other parameters fixed. We observe that changing cosmological parameters, \{$\Omega_{\textrm{m}}$, $\sigma_8$\}, tilts the FPS about a pivot scale (although these pivot scales are different), whereas changing $\Gamma_{12}$ roughly changes the amplitude.}
\label{fig:fps_params}
\end{figure*}

Here, we present a preliminary analysis to address this question. In Fig.\ref{fig:fps_params}, we show the effect on the lognormal FPS of individually changing the cosmological parameters, \{$\Omega_{\textrm{m}}$, $\sigma_8$\}, and the parameter $\Gamma_{12}$. We work at $z=2$ where the lognormal model performs best.\footnote{Unlike the spectra generated in previous cases, where we use CAMB matter power spectrum, here we use linear power spectrum of \citep{Eisenstein_1998} to generate the Gaussian density field since we require to vary cosmological parameters.} Black curves show FPS for a set of fiducial parameters where the cosmological parameters, \{$\Omega_{\textrm{m}}$, $h$, $\sigma_8$, $n_s$\}, and the astrophysical parameters, \{$T_0$, $\gamma$, $\Gamma_{12}$\} are fixed to their true values, $x_{\textrm{J}}$ is the best-fit value from 2D $\chi^2$ analysis and $\nu$ is set to unity. In red (blue) curves, we decrease (increase) the parameters by 20\% one by one, keeping other parameters fixed. We find that $\Gamma_{12}$ affects FPS in relatively simplistic way, as in decreasing (increasing) $\Gamma_{12}$ increases (decreases) power at all scales. This is fully consistent with our qualitative discussion in the previous section. The effects of \{$\Omega_{\textrm{m}}$, $\sigma_8$\}, on the other hand, are more complicated. Changing these parameters "tilts" the FPS about some scale $k$. Given the fact that the nature of this effect is very different from that of $\Gamma_{12}$, we can speculate that $\Gamma_{12}$ would be non-degenerate with the two cosmological parameters. However, we would like to emphasize that Fig.~\ref{fig:fps_params} does not show that $\Gamma_{12}$ and \{$\Omega_{\mathrm{m}}$, $\sigma_8$\} are uncorrelated (which would require a full MCMC analysis). Any correlation between them might potentially systematically bias cosmological parameters.

It will be very interesting to study the results of simultaneously varying the astrophysical parameters (which can vary with redshift) and cosmological parameters (which enter the model at each redshift in the same manner) to simultaneously model the SPH data over a broad range of redshifts such as, e.g., $2\leq z\lesssim2.5$. We leave this analysis to future work.

\subsection{Comparison with other methods}
\noindent
As stated in \citetalias{Arya_2023}, various semi-numerical methods have been developed to efficiently simulate Ly$\alpha$ forest, with potential applications in parameter space exploration. The density and velocity distributions in these models are either generated using cosmological N-body simulations or using physically motivated approximations. We have herein, briefly described few such approaches to further emphasize the strengths as well as weaknesses of lognormal model when compared against them. Few such methods include:
\begin{itemize}
    \item Assuming baryons trace dark matter in simulations \citep{petitjean_1995, croft_1998}. The simplest technique is to assume the baryonic density field perfectly traces the dark matter dark matter field. While this is a safe approximation at large scales, pressure effects cannot be ignored at small scales. Our work accounts for these pressure effects to some extent, by smoothing the dark matter density field using a Gaussian filter.

    \item Simulating few handful of full hydrodynamic simulations for parameters corresponding to a "best-guess" model and Taylor expanding the observables around those best-guess values \citep{viel_2006, walther_2021}. The method considerably reduces the computation cost by eliminating the need for running large number of hydrodynamical simulations but can be inaccurate for large displacements in parameter space and is prone to underestimating errors on recovered parameters due to unaccounted errors in approximation. The immense speed of lognormal model allows us to directly calculate the flux statistics at every point in parameter space, thereby eliminating the uncertainty induced from Taylor expansion.

    \item Running a large number of inexpensive simulations (e.g., hydro-particle-mesh (HPM)) on a parameter grid and calibrating them using a small number of full hydrodynamic simulations \citep{lymas, lymas2}. The method is shown to produce substantially more accurate clustering statistics than Fluctuating Gunn-Peterson Approximation (FGPA) but still remains computationally challenging.

    \item Using emulators based on Gaussian processes / neural network / machine learning to predict flux statistics at any given set of parameters from a small number of full hydrodynamical simulations \citep{cabayolgarcia2023neural, Sinigaglia_2022, Pedersen_2021}. E.g., In \citep{cabayolgarcia2023neural}, authors have built emulators capable of reproducing various flux statistics with accuracy upto $\lesssim 2\%$, much better than our work. Despite this, the ease of implementation of the lognormal approximation makes it a viable option.
\end{itemize}

\section{Conclusions}
\label{sec:conclude}

Efficient semi-numerical models of the Ly$\alpha$ forest are expected to play an important role in interpreting the high-quality data obtained from QSO absorption spectra surveys. The focus of this work is to build on our earlier work \citetalias{Arya_2023} and understand the effectiveness of one such model, namely, the lognormal model of baryonic densities, in recovering the thermal and ionization properties of the IGM. This is done by comparing the lognormal model with Sherwood SPH simulations \citep{bolton+17-sherwood} across redshifts $2 \leq z \leq 2.7$ and investigate the recovery of the thermal parameters $T_0$ and $\gamma$ and the photoionization rate $\Gamma_{12}$. We employ an MCMC based method to carry out the comparison, using two transmitted flux statistics: the mean flux and the flux power spectrum $P(k)$.

We find that the conventional lognormal model where the baryonic density is related to the linearly extrapolated density contrast as $n_b \propto \mathrm{e}^{\delta^L_{\mathrm{b}}}$ is unable to recover the parameters reliably. This is related to the fact that the model is a poor description of the underlying baryonic density PDF obtained from the Sherwood simulations. We address this limitation by extending the model through a scaling parameter $\nu$ such that $n_b \propto \mathrm{e}^{\nu~\delta^L_{\mathrm{b}}}$. This extension provides a better match to the SPH. In particular, the thermal parameters $T_0$ and $\gamma$ are recovered within $1-\sigma$ of the SPH values. However, the recovery of the photoionization rate $\Gamma_{12}$ is still discrepant with respect to the SPH, by $\gtrsim 3-\sigma$ for $z > 2.2$. 
We explore the reason for this discrepancy in some detail, and conclude that one requires more advanced modelling of the baryonic density PDF in order to overcome this limitation. 

This work opens up the possibility of using lognormal approximation to generate mock catalogs and calculate covariance matrices for large volume cosmological surveys such as DESI, WEAVE etc. Furthermore, the ability of lognormal approximation to generate arbitrary long skewers can allow us to study the cross-correlations in flux statistics across different redshift bins, which we have attempted in \citep{arya2023covariance}. Additionally, the fact that lognormal is able to recover thermal history with 20\% accuracy can help us set parameters for initial sampling in full hydrodynamical simulations, and / or narrow down the range of priors thus reducing significant computing time.
We also speculate that, in spite of the fact that lognormal approximation is unable to recover $\Gamma_{12}$, the model can be useful for constraining cosmological parameters. This is demonstrated by showing that the dependencies of the flux statistics, in particular $P(k)$, on the $\Gamma_{12}$ and the cosmological parameters are quite different. This indicates that one may still be able to use the model for cosmological constraints even when the recovery of $\Gamma_{12}$ is unreliable. We will explore this possibility in a future project.

\section*{Acknowledgments}

We thank R. Srianand for useful discussions.
We gratefully acknowledge use of the IUCAA High Performance Computing (HPC) facility. We thank the Sherwood simulation team for making their data publicly available. The research of AP is supported by the Associates programme of ICTP, Trieste.

\section*{Data Availability}

The Sherwood simulations are publicly available at \url{https://www.nottingham.ac.uk/astronomy/sherwood/index.php}. The data generated during this work will be made available upon reasonable request to the authors.

\bibliographystyle{JHEP}
\bibliography{references_jcap}

\label{lastpage}

\end{document}